# Analytical Drain Current Model of Nanoscale Strained-Si/SiGe MOSFETs for Analog Circuit Simulation

M. Jagadesh Kumar[1], Vivek Venkataraman[2] and Susheel Nawal[1]

[1]*Department of Electrical Engineering, Indian Institute of Technology, New Delhi 110016, India*

[2]*Department of Electrical and Computer Engineering, Cornell University, Ithaca, NY - 14850, USA*

## Abstract

*For nanoscale CMOS applications, strained-silicon devices have been receiving considerable attention owing to their potential for achieving higher performance and compatibility with conventional silicon processing. In this work, an analytical model for the output current characteristics (I-V) of nanoscale bulk strained-Si/SiGe MOSFETs, suitable for analog circuit simulation, is developed. We demonstrate significant current enhancement due to strain, even in short channel devices, attributed to the velocity overshoot effect. The accuracy of the results obtained using our analytical model is verified using two-dimensional device simulations.*

## 1. Introduction

Silicon-based MOSFETs have reached remarkable levels of performance through device scaling. However, it is becoming increasingly hard to improve device performance through traditional scaling methods. Strained-silicon devices have been receiving considerable attention owing to their potential for achieving higher performance due to improved carrier-transport properties, i.e., mobility and high-field velocity [1], and compatibility with conventional silicon processing [2]-[4]. Tremendous improvement in static and dynamic CMOS circuit performance has been demonstrated using strained SOI as well as strained-Si/SiGe MOSFETs [5].

The aim of this paper is to develop a simple current-voltage analytical model for the output current characteristics of nanoscale bulk strained-Si/SiGe MOSFETs taking into consideration (i) the effect of strain on mobility and velocity overshoot and (ii) impact ionization. The output characteristics of the strained Si/SiGe MOSFETs are calculated for different

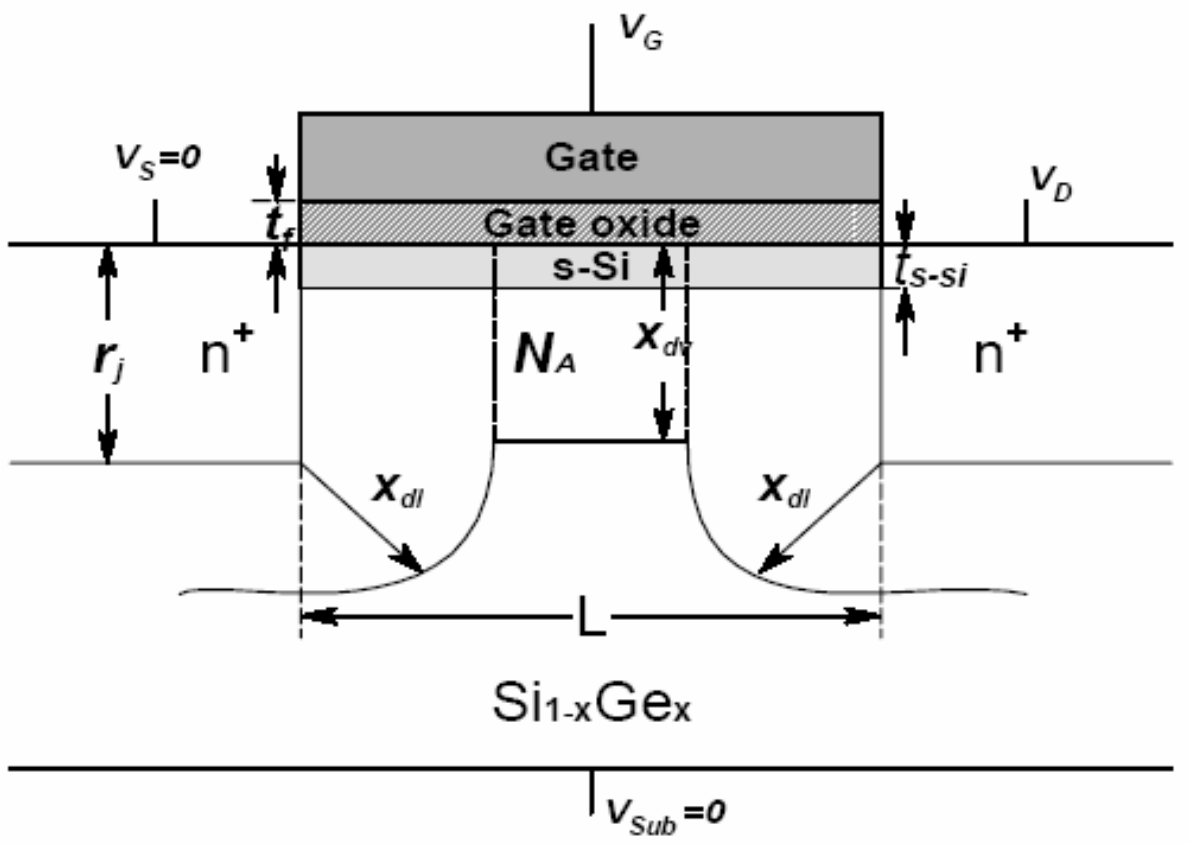

**Fig.1** Cross-sectional view of the strained-Si/SiGe MOSFET

Ge mole fractions, gate oxide thickness and gate work function. The accuracy of the model has been verified using two-dimensional simulation. It has been demonstrated that our model predicts the drain current accurately under different bias conditions.

## 2. Strained-Si/SiGe nMOSFET and the effects strain

The cross-section of the nanoscale bulk strained-Si/SiGe MOSFET considered in this study is shown in Fig. 1. The low field mobility of carriers ($\mu_{eff}$) is enhanced due to strain in Si thin films grown pseudomorphically over a relaxed SiGe substrate [6]. However, for short channel devices, high-field effects like velocity saturation work against this enhancement, and hence the benefits of strained-Si for sub-100 nm CMOS are not obvious. In spite of this, enhanced current drive and transconductance has been experimentally observed in deep submicron strained-Si devices as well [7]. Non-local effects like velocity overshoot become prominent as MOSFET dimensions shrink to the nanoscale regime, and this is directly related with the aforementioned improvement in current drive observed in short-channel MOSFETs [8].

It has been shown that an electric field step can result in the electron velocity when it exceeds the saturation velocity for a period shorter than the energy relaxation time $\tau_w$ (which is an average time constant associated with the energy scattering process, or the time needed by the electron to once again reach equilibrium with the lattice), thus causing the electron to approach ballistic transport conditions. Strain in the silicon thin film also leads to an increase in the energy relaxation time ($\tau_w$) of carriers, thus increasing the velocity overshoot [9]. Hence, to account for current enhancement in short channel strained-Si devices, the velocity overshoot effect has to be considered [10].

*A. Mobility considerations*

The low-field mobility of carriers is enhanced in strained-Si channels on SiGe substrates due to reduced phonon scattering [11] and carrier redistribution in the modified energy-subband structure [12]. The mobility enhancement factor '$e_n$' for electrons, for different values of Ge mole fraction 'x' of the relaxed SiGe substrate, is calculated based on theoretical models [11,13] as

$e_n = 1$ for $x = 0$,
$e_n = 1.46$ for $x = 0.1$,
$e_n = 1.68$ for $x = 0.2$ (1)

The above values are found to agree well with experimental data [14]. The electron mobility enhancement is found to be sustained at high values of the transverse electric field '$E_{eff}$' as well (~70% enhancement ($e_n$ = 1.7) for x = 0.2, even for $E_{eff}$ as high as 1.5 MV/cm) [3]. Using the Watt mobility model, the effective mobility of inversion layer electrons in the channel at the gate-oxide/strained-Si film interface can be written as [15, 16, 17]:

$$\mu_{eff} = \left(\mu_{ph}^{-1} + \mu_{sr}^{-1} + \mu_{c}^{-1}\right)^{-1} \quad (2)$$

where $\mu_{ph} = e_n(481)\left(\frac{E_{eff}}{10^6}\right)^{-0.16}$,

$$\mu_{sr} = e_n(591)\left(\frac{E_{eff}}{10^6}\right)^{-2.17}, \mu_c = e_n(1270)\left(\frac{N_{inv}}{10^{12}}\right)^{1.07}\left(\frac{10^{18}}{N_A}\right).$$

Here $\mu_{ph}$ is the mobility associated with phonon scattering, $\mu_{sr}$ is the mobility associated with surface roughness scattering and $\mu_c$ is the mobility associated with coulomb/ionic impurity scattering, all in units of $cm^2$/V.s. The transverse electric field $E_{eff}$ is given by [18]:

$$E_{eff} = \frac{q}{\varepsilon_{Si}}\left(N_d + \frac{N_{inv}}{2}\right) \quad (3)$$

where $N_{inv} \cong \frac{C_{ox}}{q}\left(V_{GS} - V_{th}\right)$ and $N_d \cong N_A x_d$.

In equation (3), $N_{inv}$is the inversion electron sheet density (per unit area), $N_d$ is the bulk depletion charge density per unit area under the gate, and $x_d$ is the average depletion depth under the gate as given below:

$$x_d \cong \frac{2x_{dl}\left(r_j + \frac{\pi}{4}x_{dl}\right) + \left(L - 2x_{dl}\right)x_{dv}}{L}, \quad L \geq 2x_{dl} \quad (4)$$

$$x_d \cong r_j + \frac{1}{2}\sqrt{x_{dl}^2 - \frac{L^2}{4}} + \theta\frac{x_{dl}^2}{L}, \quad L \leq 2x_{dl} \quad (5)$$

where $\theta = \sin^{-1}\left(\frac{L}{2x_{dl}}\right)$, and $x_{dl} = \sqrt{\frac{2\varepsilon_{SiGe}V_{bi,SiGe}}{qN_A}}$

is the lateral source-body and drain-body depletion region width. The parameters in equations (4) and (5) are defined in Fig. 1.

*B. Velocity overshoot effects*

To account for velocity saturation in nanoscale devices at high longitudinal electric fields, the following two-region piecewise empirical model for velocity $v_{DD}(x)$ versus longitudinal electric field $E_x$ (for electrons in the inversion layer) has been used [19, 20]:

$$v_{DD}(x) = \frac{\mu_{eff}}{1 + \frac{\mu_{eff}E_x}{2v_{sat}}}E_x \quad \text{for } E(x) \leq E_{sat} \quad (6)$$

$$v_{DD}(x) = v_{sat} \quad \text{for } E(x) > E_{sat} \quad (7)$$

where $E_{sat} = 2v_{sat}/\mu_{eff}$ is the saturation electric field, $v_{sat} = 10^7$ cm/s is the saturation velocity [15], and x is the distance from the source along the channel.

The above is a simple drift-diffusion model for carrier transport. However, for nanoscale devices, non-local effects like velocity overshoot play a significant role. This overshoot occurs in scaled devices because of the large gradient in the longitudinal electric field in the channel, and the average carrier transit time from source to drain being comparable to, or less than, the average energy relaxation time, $\tau_w$ [21]. The carrier kinetic energy, or equivalently the carrier temperature, lags the local field due to this finite energy relaxation time $\tau_w$, or relaxation length $\delta(E_x)$. When carriers are injected into the high-field region of a scaled MOSFET channel, their random thermal kinetic energy is smaller than that implied by the local field. Since the carrier mobility is inversely proportional to the carrier energy, these carriers have mobilities that are high, and therefore move with velocities higher than those implied by a local velocity-field model, i.e., they experience, on the average, quasi-ballistic flow. The average velocity of the carriers can hence be higher than the saturation velocity. Thus, including the velocity overshoot effect, the expression for carrier velocity along the channel is modified as [22-24]:

$$v(x) = v_{DD}(x)\left(1 + \frac{\delta(E_x)}{E_x}\frac{dE_x}{dx}\right) \cong v_{DD}(x)\left(1 + \frac{2v_{sat}\tau_w}{3E_x}\frac{dE_x}{dx}\right) \quad (8)$$

Due to strain, the high field transport properties of carriers in the inversion layer are also modified. Although the change in the saturation velocity with strain is expected to be small, transient transport calculations at high lateral (longitudinal) fields show a significant enhancement of the transient velocity overshoot with increasing energy splitting between the conduction subbands i.e. with increasing strain [9]. This effect can be attributed to an increase in the energy relaxation time with increasing strain [7]:

$\tau_w = 0.1ps$ for x=0, $\tau_w = 0.15ps$ for x=0.1, $\tau_w = 0.2ps$ for x=0.2 (9)

where x is the Ge mole fraction in $Si_{1-x}Ge_x$ substrate.

From equations (6) and (8), we get

$$v(x) = \frac{\mu_{eff}}{1 + \frac{\mu_{eff}E_x}{2v_{sat}}} E_x\left(1 + \frac{2v_{sat}\tau_w}{3E_x}\frac{dE_x}{dx}\right) \quad (10)$$

To estimate the gradient of the longitudinal electric field along the channel, we assume a quadratic variation of the electrostatic potential along the channel as

$$V(x) \cong \left(1 - \frac{a}{2}\right)\frac{V_{DS}}{L}x + \frac{a}{2}\frac{V_{DS}}{L^2}x^2 \quad (11)$$

where $a$ is a constant that could be dependent on device parameters and technological features of the MOSFET. Comparing model with simulation, we get $a \approx 0.2$. Thus,

$$\frac{dE_x}{dx} = \frac{d^2V(x)}{dx^2} = a\frac{V_{DS}}{L^2} \quad (12)$$

This approximation is quite valid in strong inversion conditions and similar expressions can also be found in [25, 26]. It helps us in finally obtaining a closed form analytical expression for the output drain current. Substituting (12) into (10) we get,

$$v(x) = \frac{\mu_{eff}}{1 + \frac{\mu_{eff}E_x}{2v_{sat}}}\left(E_x + k\frac{V_{DS}}{L^2}\right) \quad (13)$$

where $k = \dfrac{2av_{sat}\tau_w}{3}$.

## 3. Model for the output Current-voltage characteristics

To derive the current expression, we first write the current at any point x along the channel as

$$I_D = WQ_{inv}(x)v(x) = WC_{ox}\left(V_{GS} - V_{th} - V(x)\right)v(x) \quad (14)$$

where W is the device width. Therefore,

$$v(x) = \frac{I_D}{WC_{ox}\left(V_{GS} - V_{th} - V(x)\right)} \quad (15)$$

Putting (13) in (15) and using $E_x = \dfrac{dV(x)}{dx}$ we get:

$$I_D\left(1 + \frac{\mu_{eff}}{2v_{sat}}\frac{dV(x)}{dx}\right) = \mu_{eff}WC_{ox}\left(V_{GS} - V_{th} - V(x)\right)\left(\frac{dV(x)}{dx} + k\frac{V_{DS}}{L^2}\right) \quad (16)$$

By integrating the above equation from x= 0 to x= L and V(0) = 0 to V(L) =$V_{DS}$, we arrive at

$$I_D = \frac{\mu_{eff}WC_{ox}}{L\left(1 + \frac{\mu_{eff}V_{DS}}{2v_{sat}L}\right)}\left[\left((V_{GS} - V_{th})V_{DS} - \frac{V_{DS}^2}{2}\right)\left(1 + \frac{k}{L}\right) + \frac{ka}{12}\frac{V_{DS}^2}{L}\right]$$

for $V_{DS} \le V_{DS,sat}$ (17)

where $V_{DS,sat} = \dfrac{V_{GS} - V_{th}}{1 + \frac{V_{GS} - V_{th}}{E_{sat}L}}$

is the drain voltage at which the carriers at the drain become velocity saturated [20]. When $V_{DS}$ is greater than $V_{DS,sat}$, the velocity saturation or pinch-off point moves towards the source, away from the drain, by a distance $l_d$. The voltage difference $V_{DS}$ - $V_{DS,sat}$ appears across this distance $l_d$, where $l_d$ is the channel length modulation, given by [26]:

$$l_d = l_c \sinh^{-1}\left(\frac{V_{DS} - V_{DS,sat}}{l_cE_{sat}}\right) \quad (18)$$

where $l_c = \sqrt{\dfrac{\varepsilon_{av}x_{dv}}{2(C_{ox} + C_d)}}$, $\varepsilon_{av} \approx \dfrac{\varepsilon_{Si} + \varepsilon_{SiGe}}{2}$, $C_d = \dfrac{\varepsilon_{av}}{x_{dv}}$

and $x_{dv}$, the vertical depletion region depth due to gate bias only, is defined as

$$x_{dv} = \sqrt{\frac{2\varepsilon_{SiGe}(\phi_{th} - V_{sub})}{qN_A}}, \qquad \phi_{th} = 2\phi_{F,Si} + \Delta\phi_{s-Si},$$

$$\Delta\phi_{s-Si} = \frac{-(\Delta E_g)_{s-Si}}{q} + V_T\ln\left(\frac{N_{V,Si}}{N_{V,s-Si}}\right),$$

where $\phi_{th}$ is the minimum surface potential required for inversion [26], $r_j$ is the source/drain junction depth, L is the channel or gate length, and $V_{sub}$ is the substrate bias. $\phi_{th}$ is that value of surface potential at which the inversion electron charge density in the strained-Si device is the same as that in unstrained-Si at threshold [26] (i.e. $\Delta\phi_{s\text{-}Si} = 0$ for unstrained-Si). Hence, to obtain the current expression in the saturation region, we integrate equation (16) from x = 0 to x = L - $l_d$, and get

$$I_{Dsat} = \frac{\mu_{eff} W C_{ox}}{L\left(1 - \frac{l_d}{L} + \frac{\mu_{eff} V_{DS,sat}}{2 v_{sat} L}\right)} \left[ \left(V_{GS} - V_{th}\right) V_{DS,sat} \left(1 + \frac{k}{L} \frac{V_{DS}}{V_{DS,sat}} \left(1 - \frac{l_d}{L}\right)\right) - \frac{V_{DS,sat}^2}{2} - \frac{k}{2} \frac{V_{DS}^2}{L} \left(1 - \frac{l_d}{L}\right)^2 \left(1 - \frac{a}{6} - \frac{a}{3} \frac{l_d}{L}\right)\right]$$

$$\text{for } V_{DS} > V_{DS,sat} \qquad (19)$$

Equations (17) and (19) reduce to the familiar velocity saturation limited drift-diffusion current model given in [20] - and used in many previous works - for k = 0 i.e. no velocity overshoot.

To complete the analysis, we consider the impact ionization and avalanche multiplication of carriers in the high-field region near the drain in the saturation regime. The generation current (due to holes flowing into the substrate and electrons flowing out of the drain), can be written as [28]:

$$I_G = (M - 1) I_{D,sat} \qquad (20)$$

where the multiplication factor *(M-1)* is given by [29]

$$(M - 1) = \alpha(V_{DS} - V_{DS,sat}) \exp\left(\frac{-\beta}{V_{DS} - V_{DS,sat}}\right) \qquad (21)$$

where $\alpha$ and $\beta$ are fitting parameters [30]. We have used $\alpha$ = 0.15 ($V^{-1}$) and $\beta$ = 15.7 V (from [29]) in our model. For strained Si devices, the ionization rate increases with increasing strain, because of the reduction in the bandgap of Si – $(\Delta E_g)_{s\text{-}Si}$ - induced by the strain at the Si/SiGe heterointerface [9]. Hence the multiplication factor is modified as:

$$(M - 1) = \alpha\left(V_{DS} - V_{DS,sat}\right) \exp\left(\frac{-\beta}{V_{DS} - V_{DS,sat}}\right) \exp\left(\frac{(\Delta E_g)_{s-Si}}{q V_T}\right) \qquad (22)$$

Hence the total drain current in the saturation region can be written as

$$I_D = I_{D,sat} + I_G \quad \text{for} \quad V_{DS} > V_{DS,sat} \qquad (23)$$

## 4. Results and Discussion

The low-field mobility enhancement in strained-Si n-MOSFETs can be explained by suppressed intervalley scattering and reduced effective mass, due to the strain-induced conduction band energy splitting [1]. However, high-field and transient transport properties are expected to dominate the characteristics of deep submicron MOSFETs. Hence, hydrodynamic (HD) (energy balance) device simulations were carried out using MEDICI [15] to analyze the impact of low field mobility and high field transport on device characteristics. In hydrodynamic modeling of current transport, the strength of transient transport behavior is represented by the energy relaxation time $\tau_w$. With increasing strain, $\tau_w$ increases and almost doubles for x = 0.2, indicating that the transient electron velocity overshoot is significantly enhanced. The Watt surface mobility model is used to model the transverse-field dependent low field mobility, whereas, the high lateral-field transport is modeled with “carrier temperature based mobility” (TMPMOB) [16]. In this approach, the energy balance equation is locally solved concurrently with the drift-diffusion equation, to calculate the local mobility as a function of the local carrier temperature. The device parameters used in our simulation are given in Table 1.

Fig. 2 shows the drain current enhancement with change in strain (Ge content in SiGe) for a gate length of 50 nm and $V_{GS}$ = 0.75 V. However, for a particular technology, it is desirable to have approximately the same $V_{th}$ for various devices. To exclude the contribution of decrease in $V_{th}$ to current enhancement, we plot the normalized current ($I_{DS}(V_{GS}-V_{th})^{-1}$) versus drain voltage. We can observe that there is a significant increase in the drain current with increasing strain. This can be attributed to three main factors: (i) the increase in electron velocity overshoot due to increase in the energy relaxation time $\tau_w$, (ii) increase in low field mobility and (iii) decrease in threshold voltage $V_{th}$. Thus it is evident that strained Si provides current enhancement even for nanoscale devices.. Clearly, we can see that strained Si offers tremendous improvement in current drive. The model predictions are in close proximity with simulation data.

Figs. 3 and 4 show the output characteristics for different Ge mole fractions ‘*x*’ (0 and 0.2 respectively), and the corresponding gate oxide thicknesses $t_{ox}$, in order to keep roughly the same value of $V_{th} \sim 0.25$ V. It is observed that strained-Si MOSFETs are able to achieve the same current drive as conventional unstrained MOSFETs, even with a drastic increase in oxide thickness. There is a good agreement between model and simulation throughout the range of device parameters and bias conditions. Our model deviates slightly from the simulation results for $V_{GS}$ very close to $V_{th}$ and $V_{GS} < V_{th}$, because for these voltages our approximation for the inversion charge, $N_{inv} \cong (C_{ox}/q)\ (V_{GS}-V_{th})$, is not valid. However, as can be seen from the figures, an excellent agreement is obtained for gate voltages up to 1.0 V, which is expected to be the supply voltage for this technology (50 nm gate length). Fig. 5 shows the output characteristics for 20% Ge fraction in the SiGe substrate, with a mid-gap metal gate. We can again see that the model values track the simulation data well, thus confirming the validity of the model over different gate work functions as well.

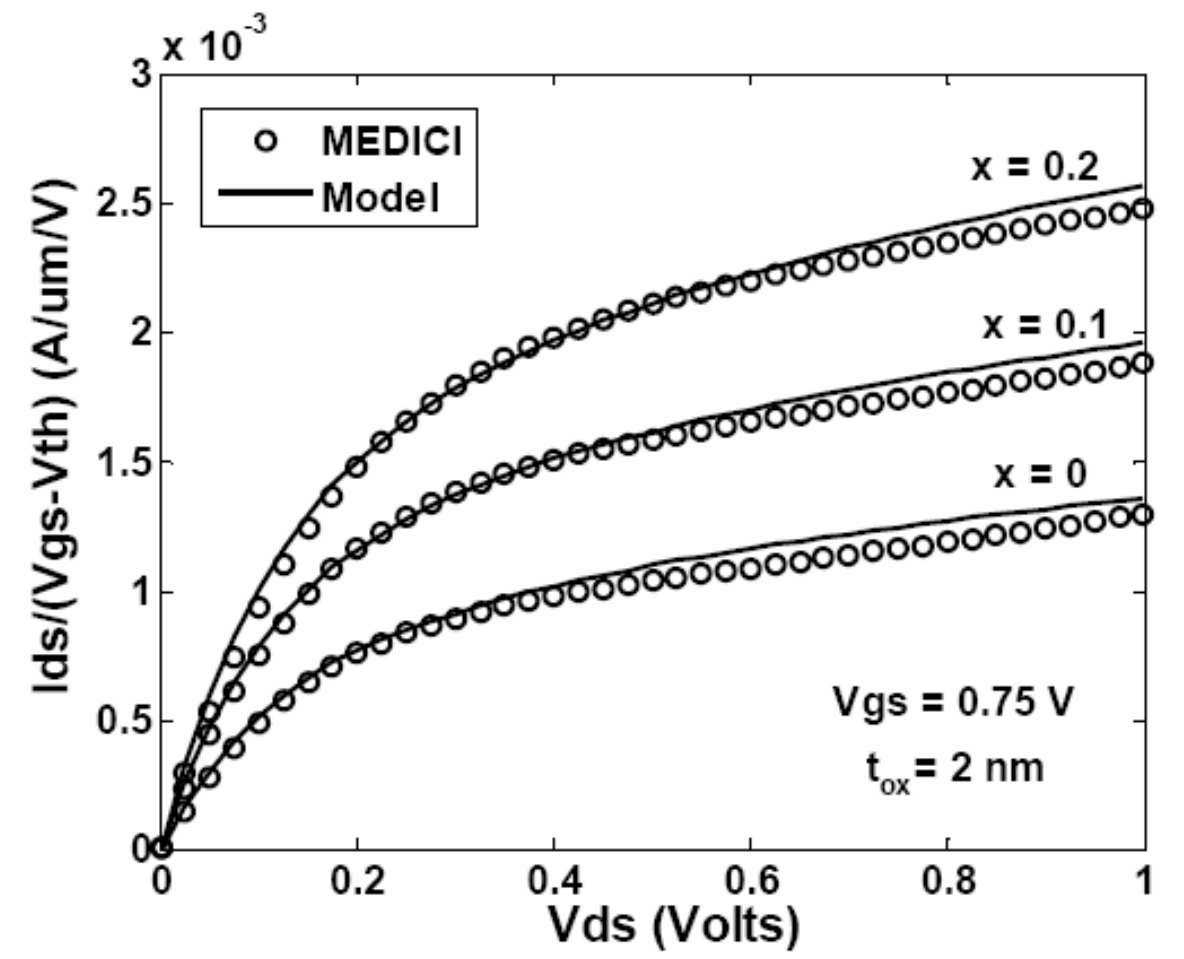

**Fig. 2** Comparison of normalized drain current for different Ge concentrations.

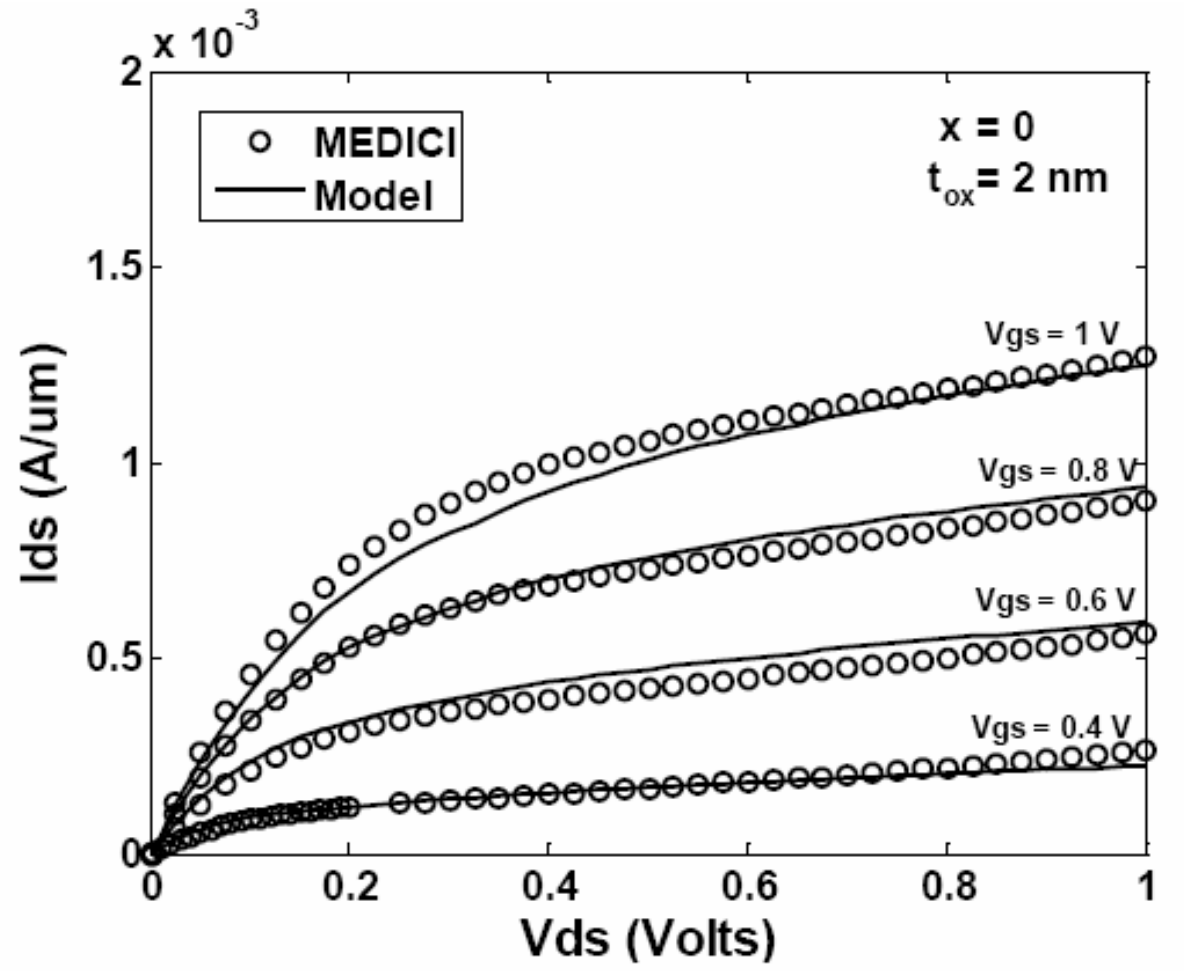

**Fig. 3** Output characteristics for conventional unstrained MOSFET ($x = 0$)

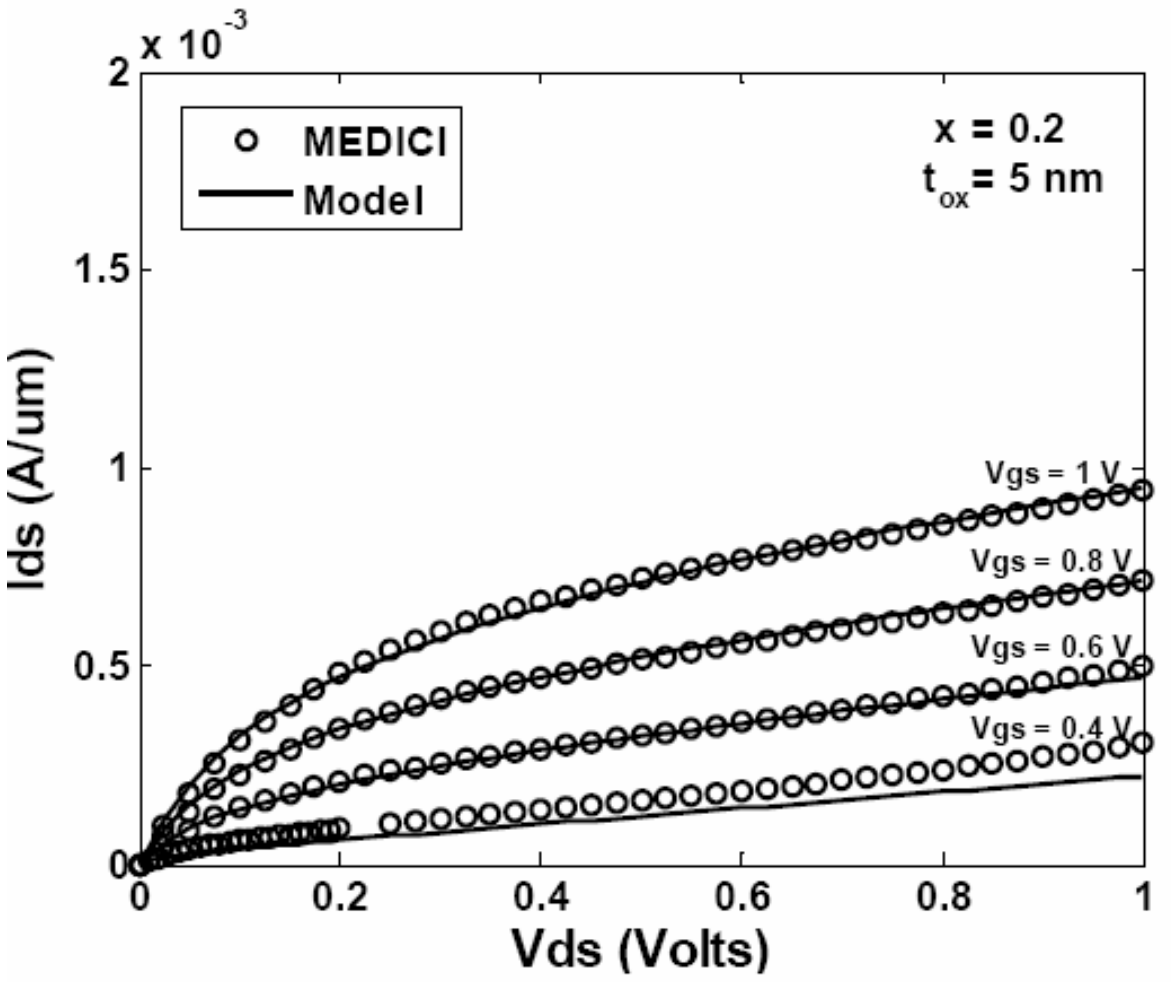

**Fig. 4** Output characteristics for $x = 0.2$ and $t_{ox} = 5$ nm

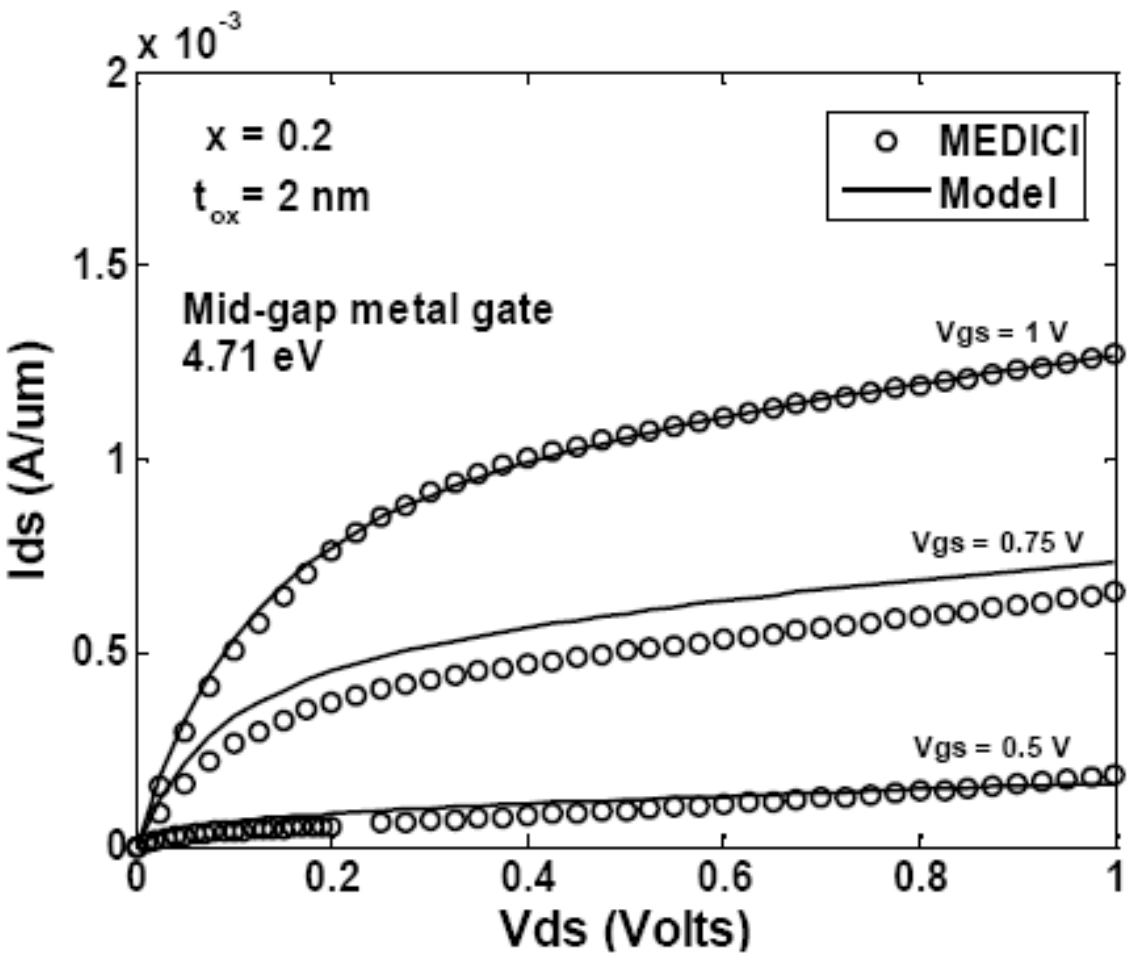

**Fig. 5** Output characteristics for $x = 0.2$, $t_{ox} = 2$nm and $\phi_M = 4.71$ eV

**Table 1:** Device parameters used in the simulation for the output characteristics of s-Si/SiGe MOSFET

| Parameter | Value |
|---|---|
| Ge mole fraction of SiGe substrate, $x$ | 0 – 0.2 (0 – 20%) |
| Source/Drain doping | $2x10^{20}$ $cm^{-3}$ |
| Body doping, $N_A$ | $10^{18}$ $cm^{-3}$ |
| Gate Length, $L$ | 50 nm |
| Gate Oxide Thickness, $t_f$ | 2.0 nm – 6.0 nm |
| Work function of gate material, $\phi_M$ | 4.35 eV (n+ poly Si) |
| Strained-Silicon film thickness, $t_{s-Si}$ | 15 nm |
| Source/drain junction depth, $r_j$ | 50 nm |
| Substrate bias, $V_{sub}$ | 0 Volts (Gnd) |
| Drain bias, $V_{DS}$ | 0.0 – 1.0 Volts |
| Gate bias, $V_{GS}$ | 0.4 – 1.0 Volts ($V_{th} \sim 0.25$ V) |

## 5. Conclusions

Strain in the Si channel is emerging as a powerful technique of increasing MOSFET performance. In this paper, we have developed a simple analytical model for the current-voltage characteristics of strained-Si/SiGe MOSFET. Our model has been verified for its accuracy using two-dimensional simulation under different bias conditions and technology parameters. Our results show that strain-induced enhancements will persist even for extremely short channel length devices. Non-equilibrium high-field effects like velocity overshoot contribute highly to the increase in current drive of these nanoscale devices. Improvements in n-MOSFET performance can be obtained in a wide range of operating conditions with moderate strain. Experimental evidence corroborating the same is also widely reported [7,31].